\documentclass[11pt,a4paper]{article}
\usepackage{amsmath}
\usepackage{graphicx} % Include figure files
\usepackage{subfigure}
\usepackage[blocks]{authblk}
\usepackage{hyperref}
\usepackage[dvipdfmx, truedimen,%
top=30truemm,bottom=30truemm,%
left=25truemm,right=25truemm]{geometry}
\begin{document}
\pagestyle{empty}

%JAEA Conference recordでは、和文目次を作成いたしますので、日本語の標題と
%著者名 (全員) を別紙に明記するようお願いいたします。
\title{\bf Calculation of Fission Fragment Yields for thermal neutron reaction of $^{239}$Pu}%日本語名：239Puの熱中性子誘起核分裂における核分裂収率の計算
\author[ ]{Futoshi Minato}%日本語名：湊太志
\affil[ ]{Department of Physics, Kyushu university, Fukuoka 819-0395, Japan}
\affil[ ]{RIKEN Nishina Center for Accelerator-Based Science, Wako, Saitama 351-0198, Japan}
\affil[ ]{Email: minato.futoshi@phys.kyushu-u.ac.jp}

\date{}

\maketitle

%%%%%%%%% main %%%%%%%%%%%%%%%%%%%%%o%%%%%%%%%%%%%%%%%%%%%%%%
\begin{abstract}
Fission fragment yield evaluation in the past has been done mainly by considering independent and cumulative fission yields.
In addition to them, the fission fragment yields are related to various observables such as total kinetic energy, prompt fission neutron, decay heats, etc.
Various improvements were carried out in the fission fragment yield evaluation of JENDL-5~\cite{JENDL5}, however it did not consider correlations between such fission observables and fragment yields.
A new evaluation including fission observables is thus required for next generation of evaluated data.
We recently developed a new calculation system using CCONE code to calculate fission fragment yields.
This calculation system enables us to compare various fission observables with the experimental data simultaneously. 
In this work, we calculated thermal-induced fission of $^{239}$Pu.
We present the result of prompt fission neutrons, decay heats, and delayed neutrons.
\end{abstract}

\section{Introduction}
Fission is a phenomenon that generates various by-products in addition to fission fragments such as prompt neutrons (PN), decay heats (DH), and delayed neutrons (DN).
The number of PN is also concerned with excitation energies of fragments, which are dependent on total kinetic energies (TKE) of fragments. 
It is desirable to prepare evaluated data that have correlations between those fission observables and fragment yields.
In JENDL-5~\cite{JENDL5} fission fragment yield data are newly evaluated, where several improvements in spin-parity states, isomers, and shell corrections have been done successfully. 
However, correlations between the fission observables and fragment yields were not considered when JENDL-5 was made.
\par
We recently developed a new calculation system~\cite{Minato2024} for fission fragment yields using CCONE~\cite{Iwamoto2016}.
Within the framework in CCONE, one can estimate fission fragment yields and fission observables and compare them with the experimental data simultaneously.
The evaluation of thermal neutron-induced fission on $^{235}$U was carried out in Ref.~\cite{Minato2024}.
We recently carried out the calculation of fission fragment yields of thermal neutron-induced fission on $^{239}$Pu.
We will present our results in this paper.
\section{Formalism}
\label{sect:II}
The calculation method for fission fragment yield of $^{239}$Pu is almost the same as that of $^{235}$U~\cite{Minato2024}.
There are some characteristic points that are different from $^{235}$U and we will illustrate them mainly in this paper.
\par
\begin{figure}
\centering
\includegraphics[width=0.48\linewidth]{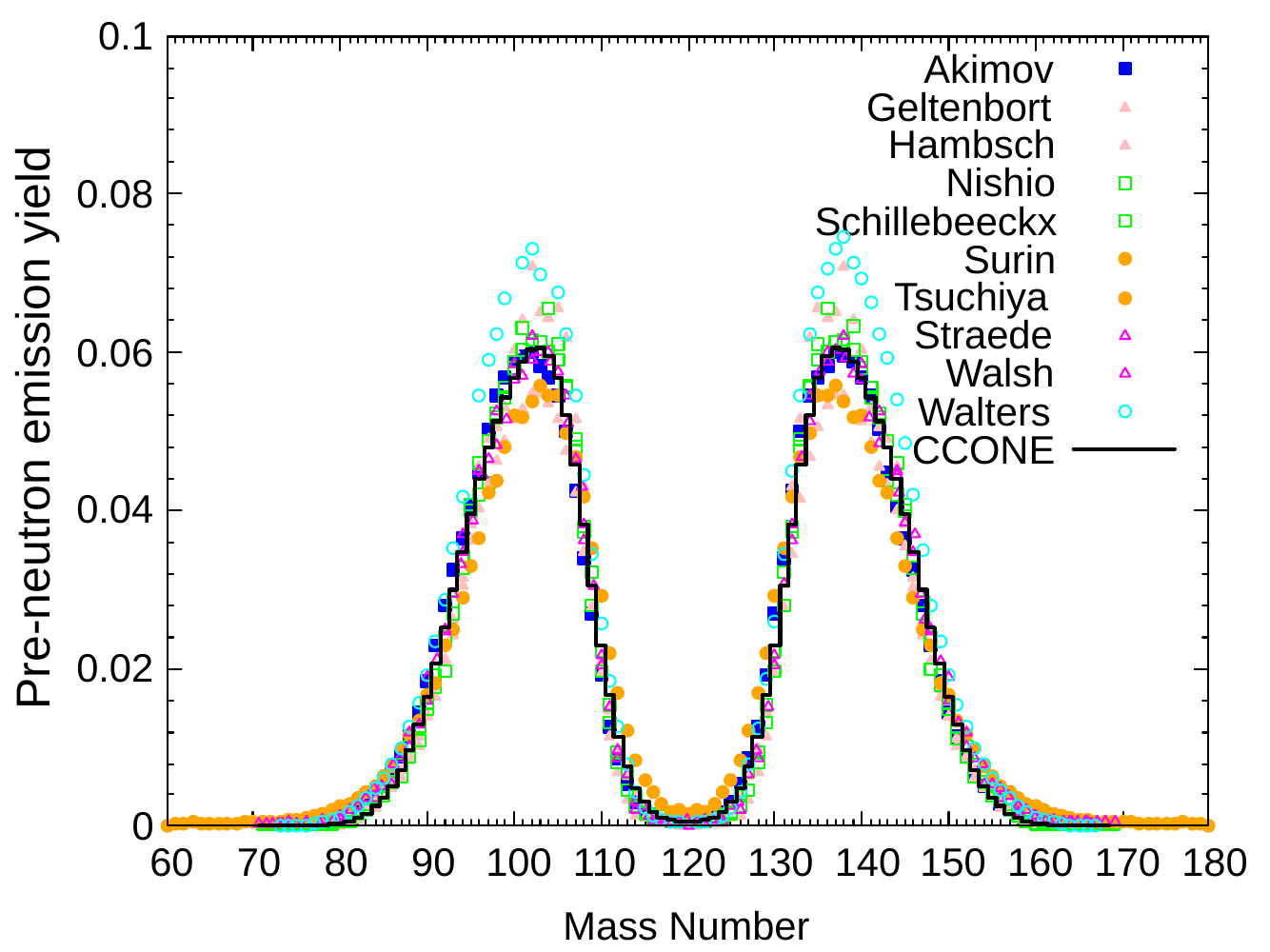}
\includegraphics[width=0.48\linewidth]{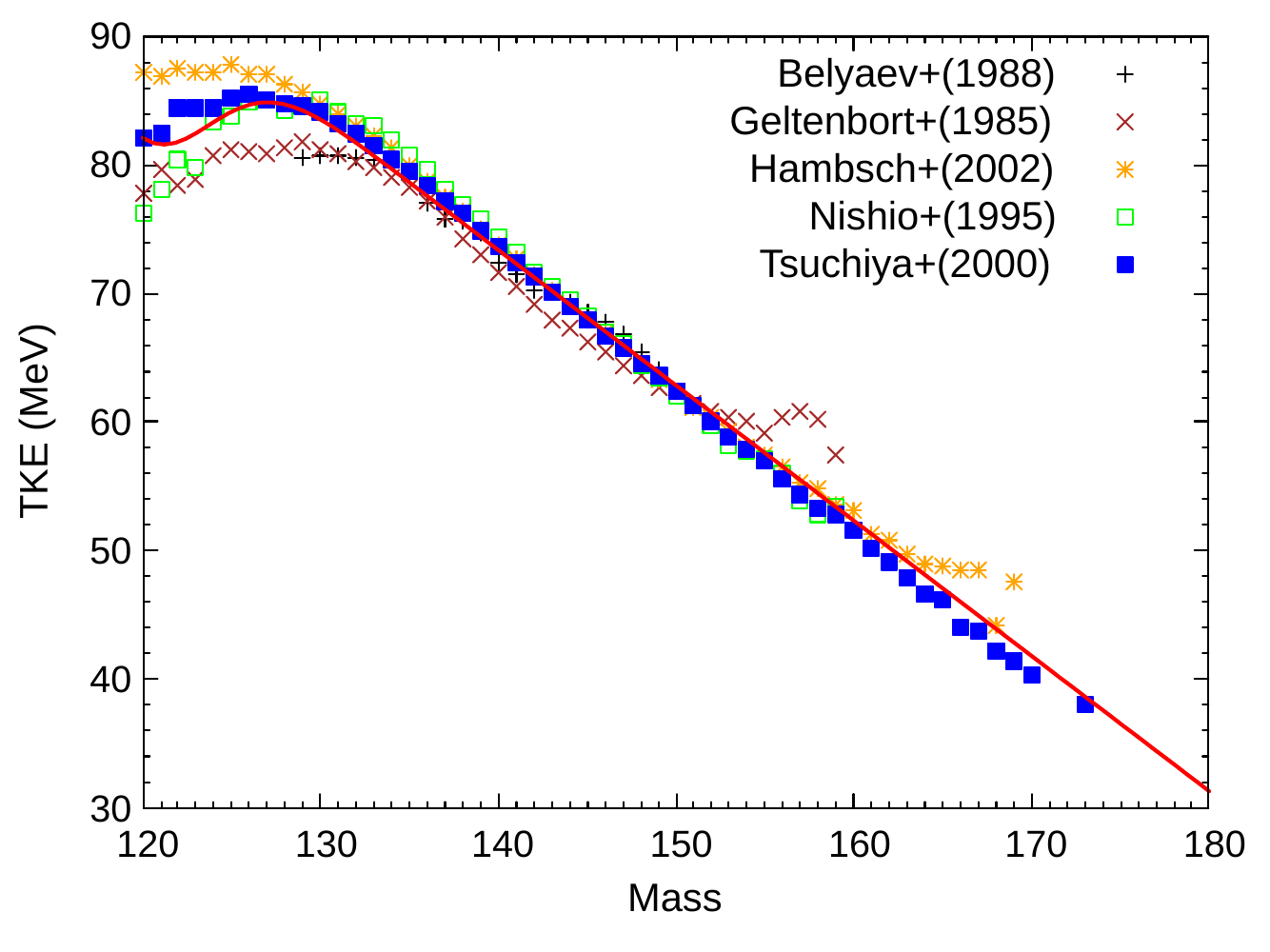}
\caption{(left) The PFY calculated from a sum of four Gaussians and experimental data. (right) The TKE curves fitted by Eq.~\eqref{eq:tke} (red solid line) and experimental data.}
\label{fig:preY}
\end{figure}

The main difference is the pre-neutron emission mass yields (PFY).
They are approximated by a sum of four Gaussians
\begin{equation}
    Y(A)=N_{1}Y_{1}(A)+N_{2}Y_{2}(A)
    \label{eq:preY}
\end{equation}
with
\begin{equation}
    Y_{i}(A)=\frac{1}{\sqrt{2\pi}\sigma_{i}}
    \left\{
    \exp\left(-\frac{(A-\mu_{i}/2)^2}{2\sigma_{i}^{2}}\right)
    +\exp\left(-\frac{(A-(A_{c}-\mu_{i}/2))^2}{2\sigma_{i}^{2}}\right)
    \right\}
    \label{eq:preY2}
\end{equation}
for $i=1,2$.
$A_{h}$ and $A_{c}$ are the mass of a heavy fragment and a compound nucleus, respectively.
The normalization constants $N_{i}$ in Eq.~\eqref{eq:preY} $\mu_{i}, \sigma_{i}$ in Eq.~\eqref{eq:preY2} are determined from experimental data (Table~\ref{table:exp}) via the least square fitting. 
We obtained $N_{1}=0.21, \mu_{1}=14.44, \sigma_{1}=3.593, N_{2}=0.79, \mu_{2}=20.60,$ and $\sigma_{2}=6.299$.
The result is shown in the left panel of Fig.~\ref{fig:preY} together with experimental data.
\begin{table}
\centering
\caption{Experimental data used for the parameter search.}
    \begin{tabular}{|c|c|}
    \hline
    \textbf{PFY} & \cite{Akimov1971}, \cite{Geltenbort1985}, \cite{Hambsch2002},
    \cite{Nishio1995}, \cite{Schillebeeckx1980}, \cite{Surin1972}, \cite{Tsuchiya2000}
    \cite{Wagemans1984}, \cite{Walsh1981}, \cite{Walter1964} \\
    \hline
    \textbf{TKE} & \cite{Belyaev1984}, \cite{Geltenbort1985}, \cite{Hambsch2002},
    \cite{Nishio1995}, \cite{Tsuchiya2000}\\
    \hline
    \textbf{IFY} & \cite{Guenther1972}, \cite{Dubey}, \cite{Srivastava1986}, 
    \cite{Naik2022}, \cite{Bogdzel1991}, \cite{Fritze1958}\\
    \hline
    \textbf{CFY} & \cite{Naik2022} \\
    \hline
    \end{tabular}
    \begin{tabular}{|c|c|}
    \hline
    \textbf{PG} & \cite{Pleasonton1972}\\
    \hline
    \textbf{DN} & \cite{Keepin1957}\\
    \hline
    \textbf{PN} & \cite{Tsuchiya2000}, \cite{Basova1979}, \cite{Fraser1966}, \cite{Batenkov2004}, \cite{Apalin1965} \\
    \hline
    \textbf{DH} & \cite{Akiyama1982}, \cite{Nguyen1997}, \cite{Dickens1980} \cite{Johanson1988} \\
    \hline
    \end{tabular}
    \label{table:exp}
\end{table}
\par
Charge distributions are approximated by the $Z_{p}$-model~\cite{Wahl2002} with modifications of the most probable atomic number $Z_{p}(A)$ and the standard deviation $\sigma_{p}(A)$ in the range of $64\le A \le 176$:
\begin{equation}
    C(A,Z)=\int_{Z-0.5}^{Z+0.5}\frac{1}{\sqrt{2\pi}\sigma_{p}(A)}\exp\left(-\frac{(Z'-Z_{p}(A))^2}{2\sigma_{p}(A)^{2}}\right).
\end{equation}
The parameters of $Z_{p}(A)$ and $\sigma_{p}(A)$ are determined from the least square fitting using the experimental data listed in Table~\ref{table:exp}.
We used independent fission yields (IFY), cumulative fission yields (CFY), the total number of prompt $\gamma$-rays per fission (PG), PN, DH, and DN for the parameter search.
\par
A total kinetic energy (TKE) is also approximated by using the following function, which is originally proposed by~\cite{Okumura2018}:
\begin{equation}
{\rm TKE}(A_{h})=\left(p_{0}-p_{1}A_{h}\right)\left(1-p_{2}\exp\left(-\left(A_{h}-\frac{A_{c}}{2}\right)^{2}p_{3}\right)\right)+\epsilon_{{\rm TKE}}.
\label{eq:tke}
\end{equation}
The parameters $p_{i}$ $(i=0,1,2,3)$ are determined so as to reproduce experimental data listed in Table~\ref{table:exp}. 
The results are $p_{0}=220.4, p_{1}=1.050, p_{2}=0.1303$, and $p_{3}=29.77$.
In Eq.~\eqref{eq:tke}, $\epsilon_{\mathrm{TKE}}$ is introduced to satisfy the average TKE value of experimental data $\overline{\mathrm{TKE}}=178$~MeV~\cite{Surin1972, Vorobeva1974, Kolosov1972}.
The result of TKE using this parameter set is shown in the right panel of Fig.~\ref{fig:preY}.
A fluctuation of TKE from Eq.~\eqref{eq:tke} is considered by introducing a width parameter
$\sigma_{\mathrm{TKE}}(A_{l})=\sigma_{\mathrm{TKE}}(A_{h})=s_{0}-s_{1}\exp\left[-s_{2}\left(A_{h}-\frac{A_{c}}{2
}\right)^{2}\right]$, where $s_{i}$ $(i=0,1,2)$ are the parameters. 
They are also determined from uncertainties of the experimental data, and we obtain $s_{0}=5.520, s_{1}=5.750, s_{3}=5512$.
%Finally, TKE distributions of fragments with $\sigma_{\mathrm{TKE}}$ are approximated by
%
%\begin{equation}
%    P_{\mathrm{TKE}}(\mathrm{TKE},A,Z)=\frac{1}{\sqrt{2\pi\sigma_{\mathrm{TKE}}^{2}}}
%    \exp\left(-\frac{(\mathrm{TKE}-\mathrm{TKE}(A,Z))^{2}}{2\sigma_{\mathrm{TKE}}^{2}(A)}\right).
%\end{equation}
%
%
\par
Excitation distributions of fission fragment yields are estimated as follows.
First, total excitation energy (${\rm TXE}$) is calculated by
${\rm TXE}=E_{n}+S_{n}+M_{f}-(M_{l}+M_{h})-{\rm TKE}$, where $E_{n}$ is the incident neutron energy, $S_{n}$ is neutron threshold of fissioning nucleus, and $M_{f}$, $M_{l}$, and $M_{h}$ are the mass of fissioning nucleus, light fragment, and heavy fragment, respectively.
To sort ${\rm TXE}$ into two fission fragments, we use an anisothermal model~\cite{Kawano2013}
\begin{equation}
    R_{T}=\frac{T_{l}}{T_{h}}
    =\sqrt{\frac{a_{h}(U_{h})U_{l}}{a_{l}(U_{l})U_{h}}},
    \label{eq:rt}
\end{equation}
where $U_{l,h}=E_{l,h}-\varDelta$ with $\varDelta$ being a correction energy attributed from pairing correlations~\cite{Mengoni1994}.
The function $a_{l,h}(U_{l,h})$ is given by $a_{l,h}(U_{l,h})=a^{*}\left(1+E_{sh}\frac{1-e^{1-\gamma U_{l,h}}}{U_{l,h}}\right)$, where $a^{*}$, $E_{sh}$, $\gamma$ are the asymptotic level density parameter, the shell correction, and the shell damping factor, respectively, and taken from Ref.~\cite{Mengoni1994}.
The mean excitation energies $E_{l,h}$ are estimated numerically from Eq.~\eqref{eq:rt} with a condition of $E_{l}+E_{h}=\mathrm{TXE}$.
The excitation energy distributions are then estimated by~\cite{Okumura2021}
\begin{equation}
    G_{l,h}(E)=\frac{1}{\sqrt{2\pi}\sigma_{l,h}}\exp\left(-\frac{(E-E_{l,h})^{2}}{2\sigma_{l,h}^{2}}\right),
\end{equation}
where $\sigma_{l,h}=\frac{E_{l,h}}{\sqrt{E_{l}^{2}+E_{h}^{2}}}\sigma_{{\rm TKE}}$.
Odd-even effects of fission fragment yields are considered via $f_{z}$ and $f_{n}$~\cite{Minato2017}.
Fission fragment yields including spin-parity distributions are then given by
\begin{equation}
Y(A,Z,J^{\pi},{\rm TKE})=\frac{1}{2}\frac{J+1/2}{2 f_{s}^{2}\sigma^{2}}
\exp \left(\frac{(J+1/2)^{2}}{2f_{s}^{2}\sigma^{2}(U)}\right) 
Y(A)C(A,Z)G_{l,h}(E)
\label{eq:yyy}
\end{equation}
The parameters for $f_{s}=2.756$ in Eq.~\eqref{eq:yyy}, odd-even effect parameters $f_{z}=1.120$, $f_{n}=1.020$ and the scaling factor of asymptotic level density parameter $f_{a}=0.6$ are adjusted by hand so that the calculated delayed neutrons and prompt neutrons come closer to the experimental data.

\section{Result}
\label{sect:result}
\begin{figure}[t]
\centering
\includegraphics[width=0.45\linewidth]{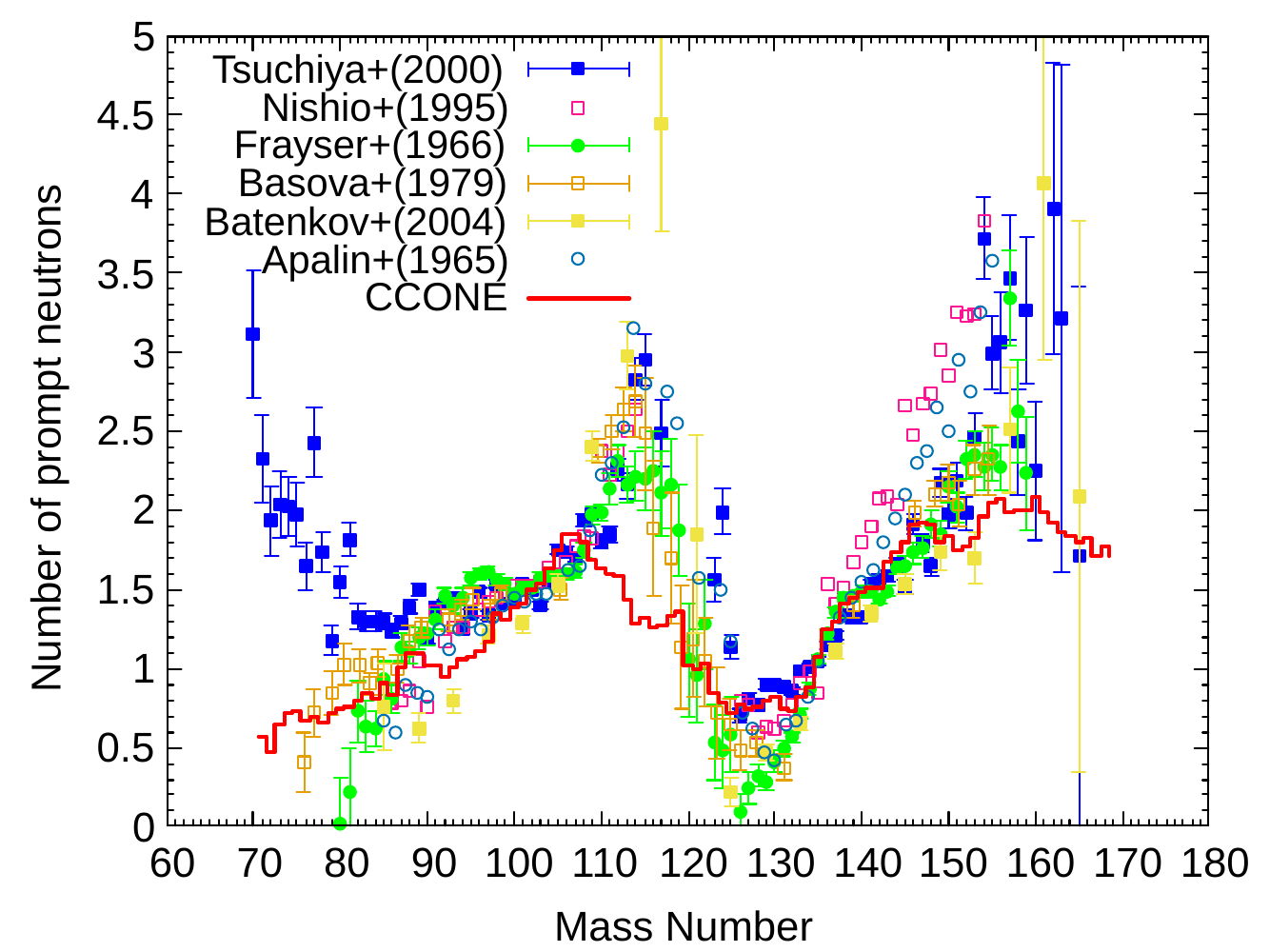}
\includegraphics[width=0.45\linewidth]{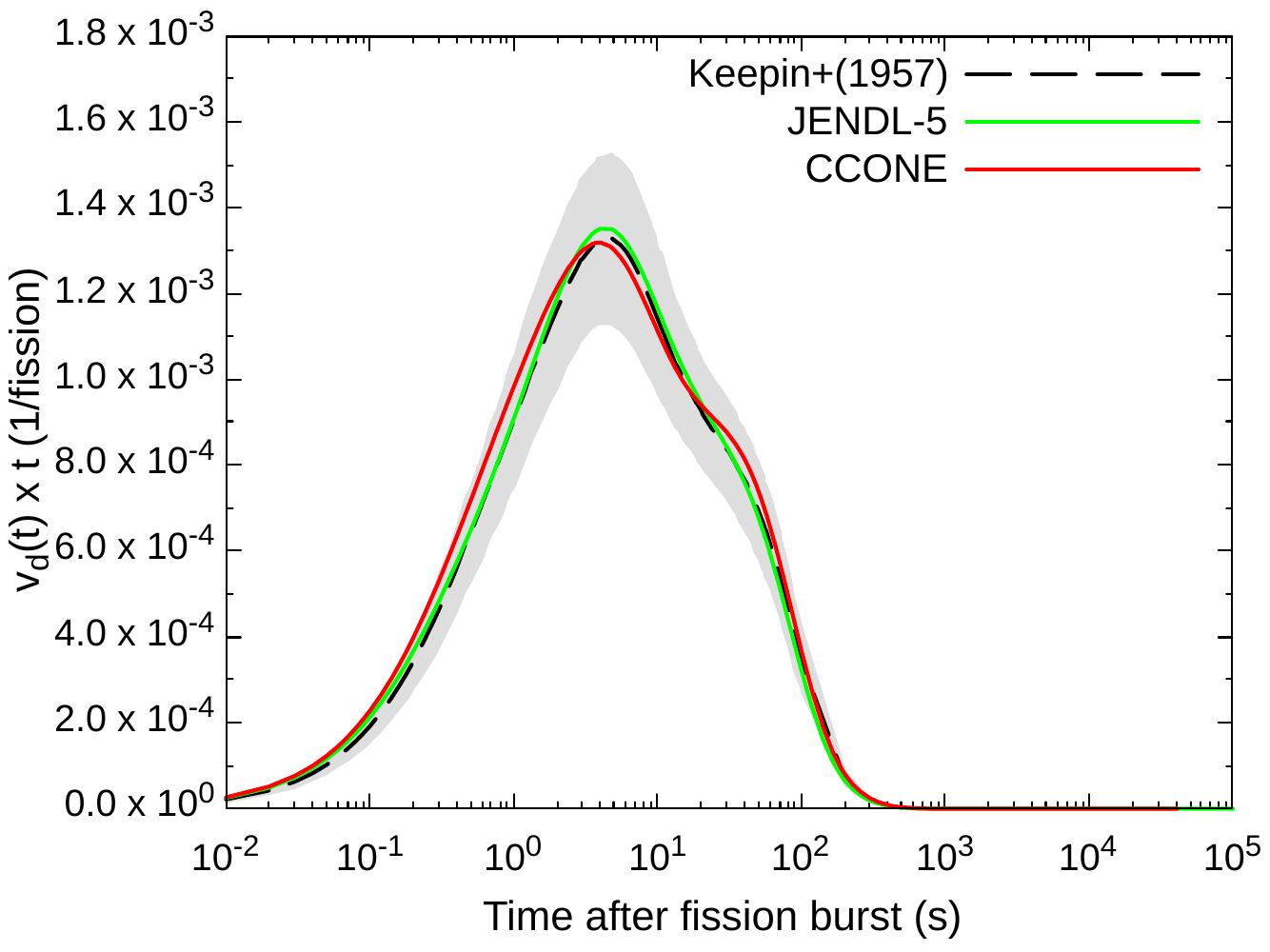}
\caption{(Left) Calculated and experimental neutron multiplicities as a function of fragment mass. (Right) Delayed neutron yields as a function of time after fission burst. The result is shown together with the JENDL-5 evaluations.}
\label{fig:result2}
\end{figure}

The result of neutron multiplicities as a function of fragment mass is shown in the left panel of Fig.~\ref{fig:result2}.
The experimental data used for the parameter search within the least square fitting are also shown.
Our calculations roughly reproduce the seesaw structure as found in the experimental data.
In this framework, the result of CCONE shows some deviations from the experimental data for light fragments ($60\le A \le 120$), while it is relatively in a good agreement with the experiments for heavy fragment sides ($135< A\le 150$).
The right panel of Fig~\ref{fig:result2} shows the delayed neutron yields as a function of time after fission burst launched by instant neutron irradiation.
The present CCONE calculation is comparable to the evaluated data of JENDL-5, which nicely reproduces the Keepin's experimental data~\cite{Keepin1957}.
\par
Figure~\ref{fig:result2} shows decay heats of $\beta$-rays and $\gamma$-rays.
The present framework of CCONE reasonably reproduces the experimental data both for $\beta$ and $\gamma$-rays.
In particular, this work improved the result of $\gamma$ decay heat as compared to the evaluation of JENDL-5 which overestimated around time after fission burst $t=80$ (s).

\begin{figure}[t]
\centering
\includegraphics[width=0.45\linewidth]{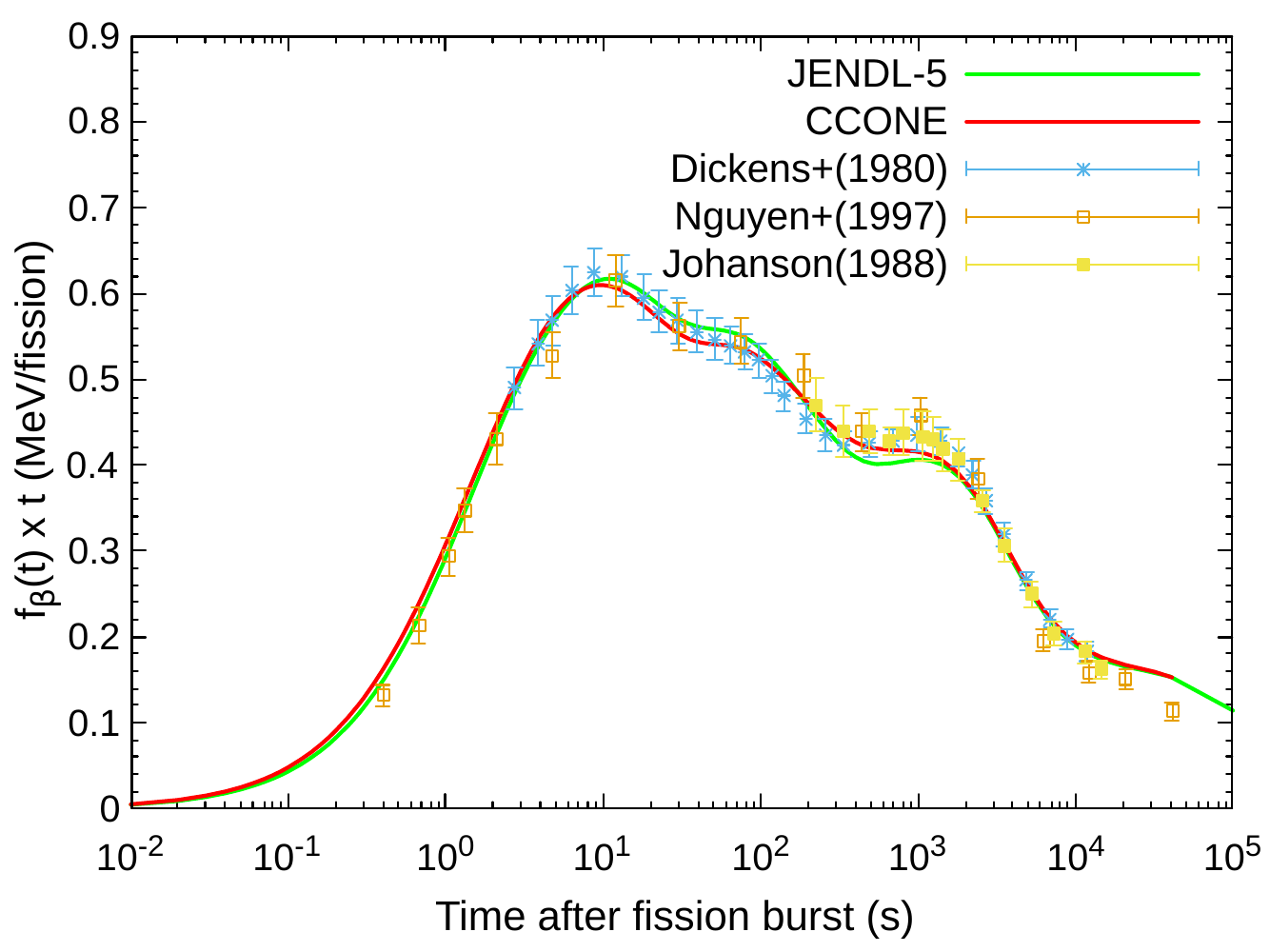}
\includegraphics[width=0.45\linewidth]{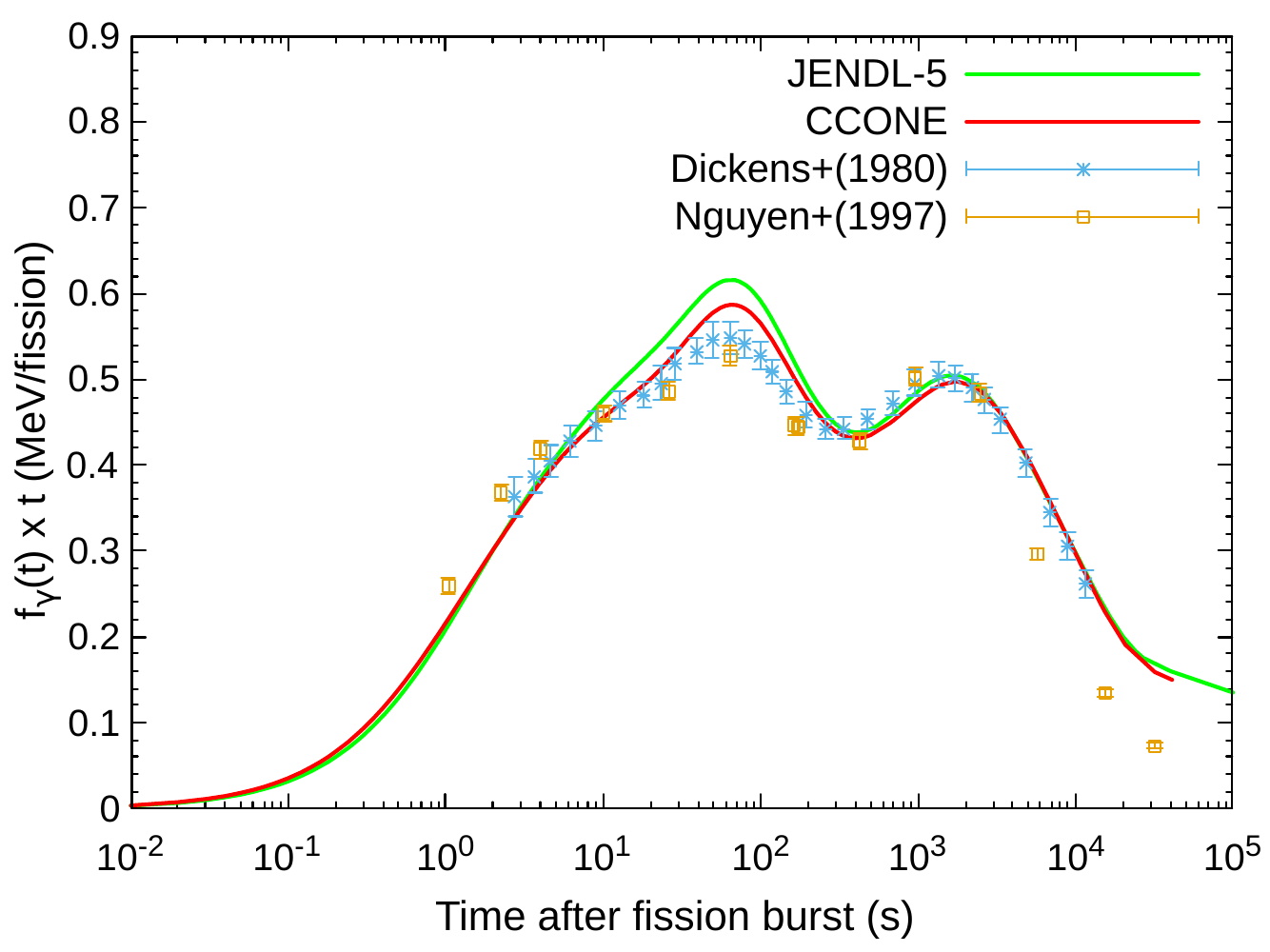}
\caption{$\beta$-ray (Left) and $\gamma$-ray (Right) decay heats as a function of fission burst.}
\label{fig:result1}
\end{figure}

\section{Summary}
We calculated fission fragment yields of thermal neutron induced fission on $^{239}$Pu with a newly developed CCONE code system.
The calculated result is compared with experimental data and the evaluation of JENDL-5.
We could obtain a reasonable agreement with experimental data and found that the result is comparable to JENDL-5 although some discrepancies from the experimental data are found in neutron multiplicities.
We will calculate fission fragment yields of different systems using this CCONE code system.

\section*{Acknowledgments}
This work is supported by JSPS KAKENHI Grant Number 21H01856 and MEXT Innovative Nuclear Research and Development Program ”Fission product yields predicted by machine learning technique at unmeasured energies and its influence on reactor physics assessment” 
entrusted to the Tokyo Institute of Technology.

\end{document}